# A novel Monte Carlo model and simulations of magnetic alloys for nuclear applications

M.Yu. Lavrentiev[*], D. Nguyen-Manh, J. Wrobel, and S.L. Dudarev

*Culham Centre for Fusion Energy, United Kingdom Atomic Energy Authority, Abingdon, OX14 3BD, UK*
[*]Corresponding Author, E-mail: Mikhail.Lavrentiev@ccfe.ac.uk

**Abstract**
We develop a Magnetic Cluster Expansion (MCE) model for binary bcc and fcc Fe–Cr alloys, as well as for fcc Fe-Ni alloys, and apply it to the investigation of magnetic properties of these alloys over a broad interval of concentrations, and over a broad interval of temperatures extending well over 1000 K. We show how an MCE-based Monte Carlo study describes the magnetic properties of these alloys, for example the composition and microstructure dependence of the Curie temperature, the non-collinearity of magnetic structures found in bcc Fe-Cr alloys, phase transitions between bcc and fcc in Fe-Cr, and the enthalpy of mixing of Fe-Ni alloys. The results of simulations are in excellent agreement with experimental observations.

## I. Introduction

Developing structural materials capable of retaining their engineering properties over extended periods of time under extreme conditions involving intense irradiation is one of the challenges for fusion and fission materials science and technology. Developing tools for efficient computer modelling of these materials is one of the objectives of the European fusion programme [1]. Here we demonstrate how the Magnetic Cluster Expansion Hamiltonian combined with Monte Carlo simulations can successfully model magnetic and thermodynamic properties of alloys that form the basis for steels developed for fusion and fission applications.

## II. Method

**1. Hamiltonian**
A Magnetic Cluster Expansion Hamiltonian was introduced in [2-4] as an effective and fast method for calculating the energy of magnetic alloys. It extends and generalizes the well known Cluster Expansion (CE) method [5] to the case of magnetic materials. While the CE method only treats the configurational disorder effects in alloys, the MCE Hamiltonian explicitly includes magnetic variables and describes the configurational alloy disorder as well as its magnetic properties. The energy of a configuration in MCE depends both on the discrete CE occupational variables and on the classical vector magnetic moments $\mathbf{M}_i$ of the constituent atoms. The moments of atoms have variable direction *and* magnitude. A complete MCE Hamiltonian contains the ordinary (non-magnetic) CE terms, the magnetic self-energy terms that determine the magnitude of magnetic moments, and inter-site Heisenberg-like magnetic interaction terms. The simplest MCE Hamiltonian has the Heisenberg-Landau form:

$$H(\{\sigma_i\},\{\mathbf{M}_i\}) = NI^{(0)} + I^{(1)}\sum_i \sigma_i + \sum_{ij} I_{ij}^{(2)}\sigma_i\sigma_j$$
$$+ \sum_i \left( A^{(0)} + A^{(1)}\sigma_i + \sigma_i \sum_j A_{ij}^{(2)}\sigma_j \right)\mathbf{M}_i^2$$
$$+ \sum_i \left( B^{(0)} + B^{(1)}\sigma_i + \sigma_i \sum_j B_{ij}^{(2)}\sigma_j \right)\mathbf{M}_i^4$$
$$+ \sum_{ij}\left( J_{ij}^{(0)} + J_{ij}^{(1)}(\sigma_i + \sigma_j) + J_{ij}^{(2)}\sigma_i\sigma_j \right)\mathbf{M}_i \cdot \mathbf{M}_j \quad (1)$$

Here, $I$'s are the non-magnetic CE coefficients, parameters $A$ and $B$ enter the configuration-dependent Landau coefficients for the magnetic self-energy terms, and $J$'s are the inter-site magnetic interaction coefficients. The functional form of the MCE Hamiltonian guarantees that the magnetic self-energy terms, and hence the directions and magnitudes of atomic magnetic moments $\mathbf{M}_i$ predicted by the model, depend on the local environment of each atom in the alloy. It is important to note that this functional form can be generalized further. In particular, one can extend the Landau expansion for the self-energy beyond the quadratic and quartic terms. This is for example necessary for modeling fcc iron. Non-magnetic alloys can also be described using the above Hamiltonian since magnetic moments vanish in equilibrium if the quadratic term coefficient is positive.

**2. Monte Carlo simulations**
Hamiltonian (1) can be readily used in Monte Carlo simulations. A large enough simulation cell (including between $10^4$ and $10^6$ atoms) can include either ordered or completely random configurations of atomic species. Then, initial vectors of magnetic moments are assigned to each atom. At each Monte Carlo step, one atom is chosen randomly and its magnetic moment is given a random change. The energy difference between the new and the old magnetic configurations is calculated and the move is accepted or declined, according to the usual Metropolis algorithm. The magnitude of change of the magnetic moment is adjusted during the simulation, so that on average half of all attempts are successful.

### III. Applications

Parameters entering the MCE Hamiltonian are derived from *ab initio* calculations performed for a number of atomic configurations, with the energy of the system and magnetic moments of each atom serving as values to be matched by the corresponding expectation values of the MCE Hamiltonian. Parameterizations were found for several systems including bcc and fcc Fe-Cr, and fcc Fe-Ni. A parameterized MCE Hamiltonian enables carrying out efficient Monte Carlo simulations for systems including up to several hundred thousand atoms.

#### 1. Magnetic properties of bcc Fe-Cr solid solutions, interfaces and clusters

We performed a combined theoretical and experimental study of magnetic properties of Fe-Cr solid solutions in the region of small Cr concentration [6]. Both experimental and computational results show that in the limit of small chromium content (up to ~6 at. %), the Curie temperature of the alloy is slightly higher than that of pure iron. Above that concentration, the Curie temperature decreases as a function of Cr content. The reason for such an unusual behaviour of the transition temperature is the strengthening of magnetic coupling in the iron subsystem in the limit of small Cr concentration due to the anti-ferromagnetic Fe-Cr coupling, whereas at higher Cr content the increased contribution of Cr-Cr interactions to the total magnetic energy diminishes the strength of ferromagnetic ordering in the alloy. Simulations involving Magnetic Cluster Expansion Hamiltonian predict the reduction of the average atomic magnetic moment as a function of Cr content, in agreement with experimental observations [6].

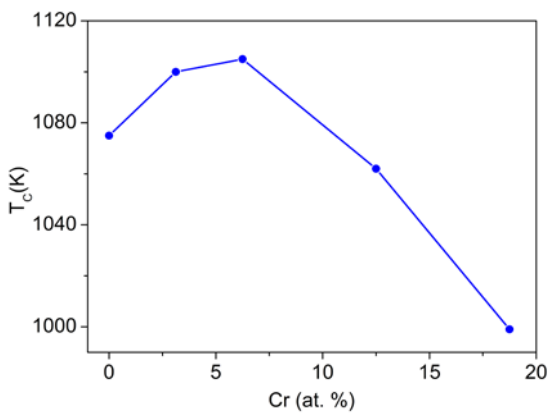

**Figure 1**: **The Curie temperature of a random Fe-Cr solid solution predicted by MCE. The occurrence of maximum of $T_C$ vs Cr content agrees with experimental observations [6].**

MCE simulations of Fe-Cr interfaces [7] showed that at a large distance from an interface, the moments of Cr atoms in the lowest energy configuration are non-collinear with respect to those of Fe atoms. A general trend exhibited by the magnetic structures follows a pattern similar to that predicted by DFT. At a large distance from an interface, we find that magnetic moments of Cr atoms are tilted with respect to, but not exactly orthogonal to, those of iron atoms. The (110) interface was found to have the highest interface energy of the three Fe/Cr interfaces modelled at low temperature. Non-collinearity of chromium magnetic moments with respect to the iron matrix was also found for Cr clusters.

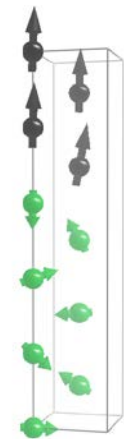

**Figure 2**: **The non-collinear magnetic structure of a Fe/Cr interface predicted by MCE. Fe atoms are black, Cr atoms are green.**

Tilting of magnetic moments at the interface observed in simulations represents a way of resolving magnetic frustration. Consider for example the (100) Fe/Cr interface with nearest-neighbour magnetic interactions favouring antiferromagnetic Fe-Cr and Cr-Cr moments alignment. If only those interactions were effective, the interfacial Cr layer would have the moment opposite to that of iron; the second Cr layer would have its moment opposite to that of the first Cr layer, etc. In this case, collinear ferromagnetic Fe and antiferromagnetic Cr structures would coexist at the interface. However, as the range of magnetic interactions extends further, the system becomes frustrated. Both the second-nearest-neighbour Fe-Cr and Cr-Cr interactions are antiferromagnetic. As a result, the second chromium layer would tend to align antiferromagnetically with respect to both Fe and the first Cr layer; as this is impossible, some adjustment of the direction of moments in both layers is necessary to reduce the total energy. Moments in the third Cr layer would then tend to align antiferromagnetically with respect to the first two, which is also impossible, resulting in the further change of the direction of moments.

Similar non-collinearity was found also in the case of chromium clusters in iron. For a large random cluster, the magnetic moments of atoms in the bulk of Cr cluster are not collinear and are in fact nearly orthogonal to the moments of iron atoms [8]. The same reasons as in the case of Fe/Cr interfaces can explain the emergence of non-collinearity here as well. Chromium clusters having regular shape have also been investigated. In the case of relatively small cubic clusters, it was found that the magnetic moments of Cr atoms largely remain collinear with respect to the Fe environment. Some noncollinearity can be observed with increasing size of clusters. The emergence of noncollinearity is demonstrated in Figure 3, where two components of the magnetic moment of Cr atoms are plotted – parallel and orthogonal to the bulk Fe atoms for the three smallest cubic clusters. For a 9-atom cluster, moments are almost completely collinear to the Fe subsystem, the largest deviation from noncollinearity is 5.5°. This angle increases to 37° for 35-atomic cluster and to 67° for the 91-atomic cluster.

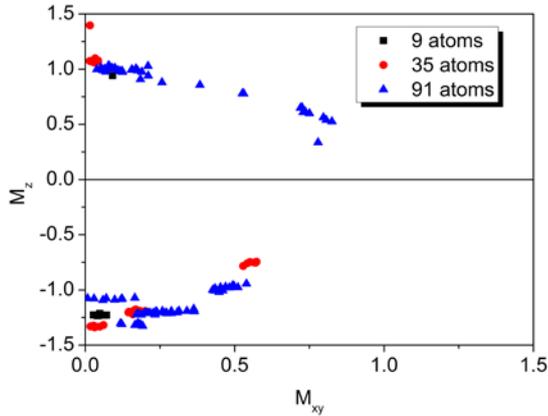

**Figure 3**: **The parallel and orthogonal, with respect to the Fe moment, components of Cr magnetic moments in the three smallest cubic clusters.**

**2. Magnetism-driven phase transitions in Fe and Fe-Cr**
Parameterization of the fcc Fe-Cr MCE Hamiltonian enables comparing the free energies of bcc and fcc Fe-Cr and calculating the phase diagram of the alloy [3]. An MCE model combined with experimentally observed information about elastic moduli, was able to describe, at a quantitative level of accuracy, various magnetic and structural phase transitions occurring in Fe-Cr magnetic alloys. It appears possible, by relying only on the zero-temperature *ab initio* data, to evaluate parameters describing thermal magnetic fluctuations in the alloy. The model correctly predicts the occurrence of two structural phase transitions below the melting point and above the Curie temperature for pure Fe and for a range of Fe-Cr alloy compositions. In Figure 4 we show that it is necessary to include all the contributions to the free energy difference between the fcc and bcc phases, including the entropy and the phonon terms, to describe the experimentally observed bcc-fcc-bcc sequence of phase transitions. It is important to mention the agreement between the MCE model and independently developed theoretical model by Körmann et al. [9-10].

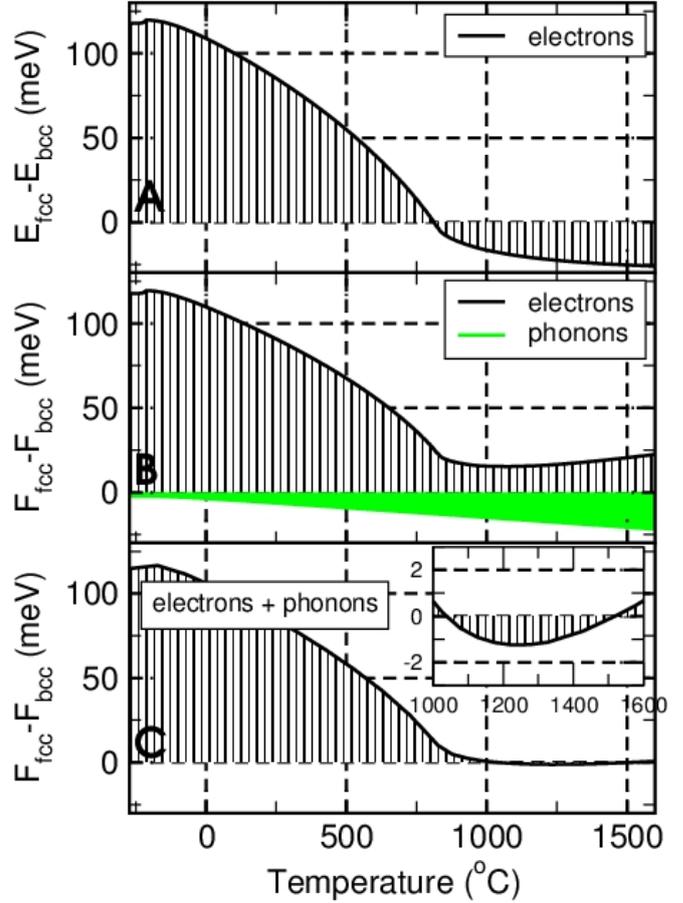

**Figure 4:** **A: fcc-bcc electron energy difference. B: fcc-bcc electron free energy difference and phonon free energy difference, shown separately. C: fcc-bcc total free energy difference, with a magnified part of the curve, illustrating the occurrence of a sequence of bcc-fcc-bcc phase transitions, shown in the inset.**

**3. Magnetism in fcc Fe-Ni**
Fcc Fe-Ni-Cr based steels are a potentially significant class of material for fusion reactors. The success of application of MCE to Fe-Cr simulations has stimulated the development of an MCE parameterization for the Fe-Ni alloy as a first step towards the treatment of the ternary Fe-Ni-Cr system. The phase diagram of the Fe-Ni alloy shows very low solubility of Ni in bcc iron, as opposite to the fcc system, where the two species are solvable in a whole range of concentrations at high temperature. A realistic MCE Hamiltonian should describe the miscibility gap between fcc Fe-Ni and the bcc Fe, as well as confirm the experimentally observed ordered $L1_2$ structure with $FeNi_3$ content. Also, the possible existence of the less stable FeNi and $Fe_3Ni$ structures (first of them was found in meteorites, see [11-12]) should be explored.

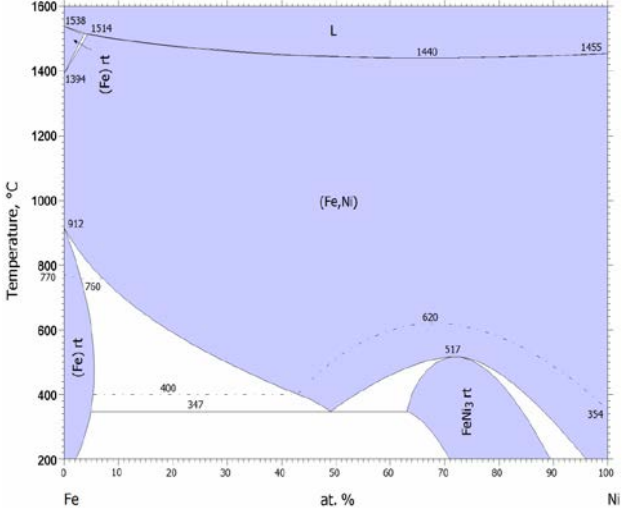

**Figure 5**: **Experimental phase diagram of the Fe-Ni. Dashed line shows magnetic order-disorder transitions. Picture adapted from [13].**

In order to parameterize the fcc Fe-Ni system, 29 configurations were studied by the *ab initio* DFT calculations, together with pure fcc iron and nickel. An important property of fcc iron is that *ab initio* studies show the occurrence of two, the so-called high-spin and low-spin, magnetic states. The high-spin configuration is slightly more energetically favourable than the low-spin one, but the difference is of the order of 10 meV/atom or even less, meaning that both configurations are important for describing the magnetic properties of the alloy. Such complex dependence cannot be accounted for by a relatively simple Landau-type Hamiltonian with only quadratic and quartic terms, which only has one minimum as function of the magnitude of the magnetic moment vector. Hence, we extend the Landau expansion up to the 8$^{th}$ order terms. This expansion can satisfactorily describe both low- and high-spin configurations of fcc Fe, as shown in Figure 6.

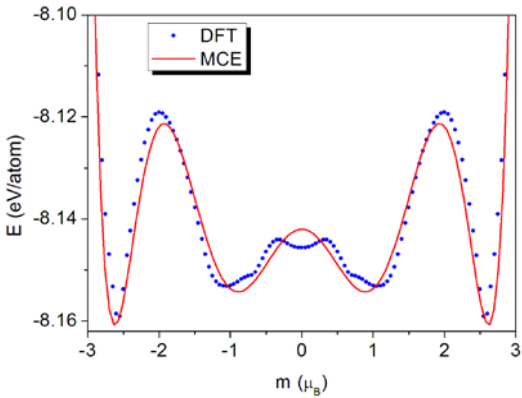

**Figure 6**: **Predicted DFT and MCE variation of energy of ferromagnetic fcc Fe as a function of magnetic moment.**

Subsequently, the interaction parameters $J_{ij}$ were fitted for Fe-Fe, Fe-Ni, and Ni-Ni magnetic interactions. Fitting was performed as follows. First, the MCE energy for each DFT structure was calculated. Then the sum of squares of energy differences between the DFT and MCE values:

$$S = \sum_i C_i \left( E_i^{DFT} - E_i^{MCE} \right) \quad (2)$$

was chosen as a measure of difference between the energies predicted by the two approaches. Here, coefficients $C_i$ were chosen so as to increase the relative weight of more important configurations – those with low energies or with content close to pure metals. Small random changes of interaction parameters were explored and then accepted or declined, depending on whether the changes decreased the value of $S$ or not. In another approach, Metropolis-like algorithm was used to decide about whether to accept or reject a change in $J_{ij}$, with the effective "temperature" decreasing in the course of the fitting procedure. Using interactions up to the 4$^{th}$ nearest neighbour, we were able to reduce the mean square deviation of MCE results from DFT down to 12 meV/atom. This value indicates very good agreement for the case of Fe-Ni, with the enthalpy of mixing being as low as -160 meV/atom for some configurations. The resulting set of parameters is given in Table 1.

|        | Fe-Fe   | Fe-Ni  | Ni-Ni   |
|--------|---------|--------|---------|
| 1st NN | -0.793  | 1.516  | -13.153 |
| 2nd NN | -10.827 | -2.710 | 7.228   |
| 3rd NN | 0.547   | -2.500 | -5.605  |
| 4th NN | 2.306   | 1.649  | -6.744  |

**Table 1**: **Magnetic interactions (in meV) fitted for the fcc Fe-Ni MCE Hamiltonian. Negative sign of a parameter corresponds to the ferromagnetic interaction.**

Using the parameterized Hamiltonian we were able to perform Monte Carlo simulations of magnetic properties of ordered and disordered Fe-Ni systems. The available experimental data on the enthalpy of mixing are rather scarce. All of them, together with several theoretical estimates, show negative enthalpy of mixing, with the absolute value of the minimum strongly varying between different approaches (see, e.g., [14], Figure 6). Our simulations confirm negative enthalpy of mixing for the random Fe-Ni system, as shown in Figure 7. The absolute value of minimum is close to -100 meV/atom, which is higher than what is found in Ref. [14]. However, this value is in agreement with our DFT calculations, which give even lower enthalpies of mixing for several ordered structures. With increasing temperature the absolute value of the enthalpy of mixing slightly increases, as shown in Figure 7.

Having accomplished the parameterization step, the equilibrium phase diagram calculation can be performed as follows. In the approximation of complete insolubility of Ni in bcc iron, the free energy of pure bcc Fe should be compared with the free energy of fcc Fe-Ni random mixture. This can be done using the well known tangent construction, see Figure 8. Curves in Figure 8 represent the free energy differences between fcc and bcc phases, $\Delta F = F_{fcc}(Fe_{1-x}Ni_x) - F_{bcc}(Fe)$. Here the free energy of $Fe_{1-x}Ni_x$ is obtained as a sum of its magnetic energy and the mixing entropy term. Then, the straight lines tangential to the curves determine the content of fcc Fe-Ni solution in equilibrium with the bcc Fe at different temperatures. As the temperature increases, the range of concentrations where the fcc system is more stable increases, in agreement with experiment (see Figure 5).

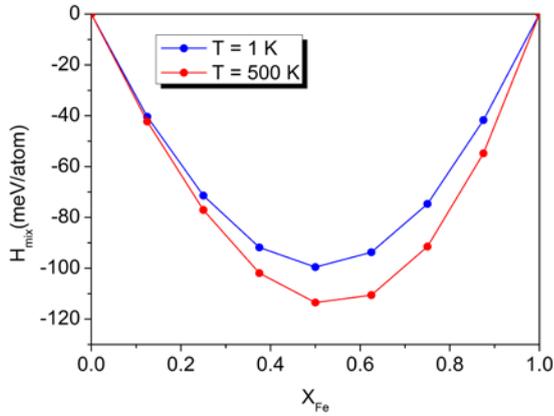

**Figure 7**: **Predicted MCE enthalpy of mixing for random Fe-Ni alloys shown as a function of alloy composition.**

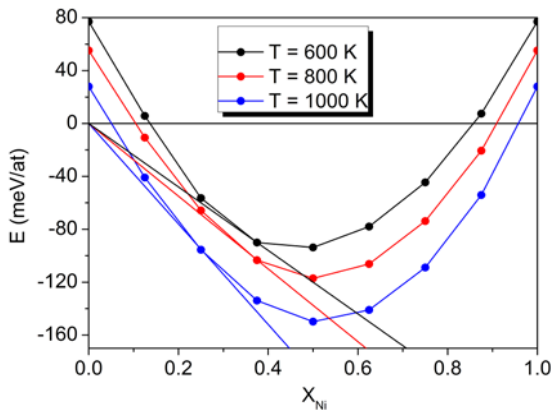

**Figure 8**: **Tangent constructions determining phase equilibrium between bcc Fe and fcc Fe-Ni.**

On the high-Ni part of the phase diagram, comparison between the ordered and random FeNi and FeNi$_3$ structures was performed. It was found that the ordered intermetallic compounds are more stable compared to the random mixtures up to the temperatures of 500 K for the FeNi and 600 K for FeNi$_3$. These structures also keep magnetic order until much higher temperatures. This leads to a maximum in the Curie temperature dependence on the Ni concentration for the FeNi$_3$ compound, corresponding to experimental data.

The resulting theoretical phase diagram for the Fe-Ni system is shown in Figure 9. All the main features of the experimental phase diagram (see Figure 5) are successfully reproduced. Agreement with experiment can be further improved by taking into account the non-zero solubility of Ni in bcc iron. This can clarify a question of whether the FeNi ordered intermetallic is stable at some temperatures, or it is metastable with respect to random fcc Fe-Ni mixture. Another important question that remains to be resolved is the existence of magnetic order at temperatures above the stability limits of ordered intermetallics. This is possible if there remains some structural order at temperatures slightly above the stability limits of ordered compounds.

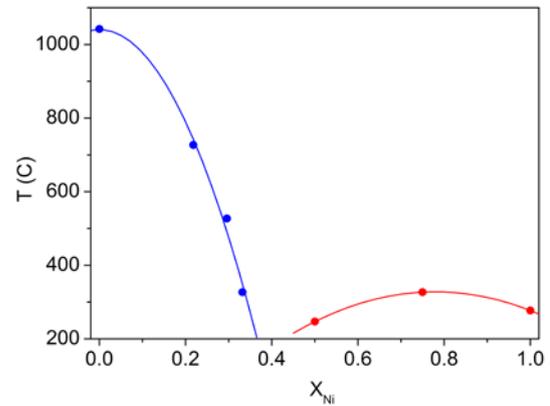

**Figure 9**: **Theoretical phase diagram of the Fe-Ni. Blue line corresponds to equilibrium between bcc and fcc Fe-Ni; red line shows stability limits for ordered Fe-Ni compounds and the magnetic order limits for them and as well as pure Ni.**

## IV. Conclusion

In summary, here we present applications of Magnetic Cluster Expansion to several systems important for fusion and fission materials science and technology. Systems such as Fe-Cr and Fe-Ni were successfully modelled using the MCE Hamiltonians parameterized on the basis of DFT calculations. We show that an MCE-based Monte Carlo study can efficiently model complex magnetic properties of Fe-based alloys, avoiding the need for time- and resources-consuming experiments. In all the cases where experimental data are available, the results derived from Monte Carlo simulation are in good agreement with experiment, as well as with *ab initio* studies.

## Acknowledgement

This work, part-funded by the European Communities under the contract of Association between EURATOM and CCFE, was carried out within the framework of the European Fusion Development Agreement. To obtain further information on the data and models underlying this paper please contact *PublicationManager@ccfe.ac.uk*. The views and opinions expressed herein do not necessarily reflect those of the European Commission. This work was also part-funded by the RCUK Energy Programme under grant EP/I501045. DNM would like to thank Juelich supercomputer centre for using the High-Performances Computer for Fusion (HPC-FF) facilities as well as the International Fusion Energy Research Centre (IFERC) for using the supercomputer (Helios) at Computational Simulation Centre (CSC) in Rokkasho (Japan). JW is supported by European Accelerated Metallurgy project.

**References**

1) S.L. Dudarev et al., *J. Nucl. Mater.* **386-388**, 1 (2009).
2) M.Yu. Lavrentiev, S.L. Dudarev and D. Nguyen-Manh, *J. Nucl. Mat.* **386-388**, 22 (2009).
3) M.Yu. Lavrentiev, D. Nguyen-Manh and S.L. Dudarev, *Phys. Rev. B* **81**, 184202 (2010).
4) M.Yu. Lavrentiev, D. Nguyen-Manh and S.L. Dudarev, *Comp. Mat. Sci.* **49**, S199 (2010).
5) J.M. Sanchez, F. Ducastelle and D. Gratias: *Physica A* **128**, 334 (1984).
6) M.Yu. Lavrentiev et al., *J. Phys. : Condens. Matter* **24**, 326001 (2012).
7) M.Yu. Lavrentiev et al., *Phys. Rev. B* **84**, 144203 (2011).
8) M.Yu. Lavrentiev, D. Nguyen-Manh and S.L. Dudarev, *Solid State Phenomena* **172-174**, 1002 (2011).
9) F. Körmann, PhD Thesis, University of Paderborn (2011), http://digital.ub.uni-paderborn.de/hs/download/pdf/8865.
10) F. Körmann et al., *Physica Status solidi (b)*, submitted.
11) C.-W. Yang, D.B. Williams and J.I. Goldstein, *Geochimica et Cosmochimica Acta* **61**, 2943 (1997).
12) R.A. Howald, *Metallurgical and Materials Transactions* **34A**, 1759 (2003).
13) I. Ohnuma, R. Kainuma and K. Ishida, in: *CALPHAD and Alloy Thermodynamics* as held at the 2002 TMS Annual Meeting, 61 (2002).
14) R. Idczak, R. Konieczny and J. Chojcan, *Physica B* **407**, 235 (2012).